\def\ANON{0} 

\documentclass[runningheads]{llncs}
\usepackage{graphicx}
\usepackage{hyperref}
\usepackage{booktabs} 
\usepackage[ruled,vlined]{algorithm2e}
\usepackage{subcaption}

\usepackage{xspace}
\usepackage{xcolor}
\definecolor{orange9}{HTML}{FFDD00}
\usepackage[color=orange9]{todonotes}

\begin{document}

\title{Time Matters: 
Exploring the Effects of Urgency and Reaction Speed in Automated Traders
\thanks{This is the authors' accepted manuscript to be published in A. P. Rocha et al. (Eds.): ICAART 2020, LNAI 12613, 2021: Springer Nature Switzerland. The final authenticated version is available online at \url{https://doi.org/10.1007/978-3-030-71158-0_7}}}

\titlerunning{Exploring Urgency and Reaction Time in Automated Traders}
%
\ifnum\ANON=0 
{
    \author{
    Henry Hanifan \and
    Ben Watson \and
    John Cartlidge \and
    Dave Cliff 
    }
    \authorrunning{H. Hanifan et al.}
    %
    \institute{Department of Computer Science, University of Bristol, Bristol, UK\\
    \email{\{hh15092,bw15485\}@my.bristol.ac.uk\\ 
    \{john.cartlidge,csdtc\}@bristol.ac.uk}}
}
\else
{
    \author{
    {\bf [ANON AUTHORS]}}
    \authorrunning{{\bf [ANON]} et al.}
    %
    \institute{{\bf [ANON INSTITUTE]}\\
    \email{{\bf [ANON EMAILS]}
    }}
}
\fi

\maketitle              
\begin{abstract}
We consider issues of time in automated trading strategies in simulated financial markets containing a single exchange with public limit order book and continuous double auction matching. In particular, we explore two effects: (i) {\em reaction speed} - the time taken for trading strategies to calculate a response to market events; and (ii) {\em trading urgency} - the sensitivity of trading strategies to approaching deadlines. Much of the literature on trading agents focuses on optimising pricing strategies only and ignores the effects of time, while real-world markets continue to experience a {\em race to zero} latency, as automated trading systems compete to quickly access information and act in the market ahead of others. We demonstrate that modelling reaction speed can significantly alter previously published results, with simple strategies such as SHVR outperforming more complex adaptive algorithms such as AA. We also show that adding a {\em pace} parameter to ZIP traders (ZIP-Pace, or ZIPP) can create a sense of urgency that significantly improves profitability. 


\keywords{Agent Based Modelling \and Auctions \and Automated Trading \and Financial Markets \and Simulation \and Trading Agents.}

\end{abstract}
%
%
%


\section{\uppercase{Introduction}}
\label{sec:introduction}

The academic literature on financial trading agents is predominately focused on strategies to determine the price at which an agent should submit the next order (often, the price most likely to maximise profit), given current market conditions. These {\em pricing strategies} are usually developed and tested in a controlled experimental environment that has changed little in form since the experimental design was first introduced in 1955 by the seminal work of Vernon L. Smith \cite{Smith62}. Smith---now regarded as the ``father of experimental economics''---borrowed techniques from psychology laboratory experiments to perform a series of trading experiments, using student volunteers at Purdue University. Smith created virtual markets by designating half the participants as buyers and half the participants as sellers. Buyers and sellers were given a range of limit prices (a private value associated with each assignment to trade) and participants interacted and negotiated freely via a simple open outcry mechanism until trade prices converged towards the market equilibrium.

In subsequent years, the majority of work on trading agents has followed Smith's deliberately simple experimental design; and with good reason, as doing so enables strict performance comparisons with earlier work. Each time a new agent design is introduced
(e.g., ZIC \cite{GodeSunder93}; ZIP \cite{Cliff97}; GD \cite{GD98}; AA \cite{Vytelingum06}), Smith's framework is used to demonstrate the relative profitability of the new agent and to measure the equilibration behaviours of markets containing the new agent, and markets containing mixtures of agent types. 
Over the decades, extensions to Smith's framework have been made, e.g., the use of limit order books \cite{DasEtal01}; real-time markets with humans and agents \cite{CartlidgeCliff12,DasEtal01,DeLucaCliff11,DeLuca11}; and more realistic market dynamics such as continuous replenishment of assignments \cite{CartlidgeCliff12,DeLuca11}, and continuously varying equilibria \cite{Cliff19,SnashallCliff19}. 
However, these extensions remain focused on studying pricing strategies, and rarely address issues of strategy timing: how long a strategy takes to compute (i.e., the {\em reaction time}), and how strategies adapt prices when there are constraints on time available (i.e., the trading {\em urgency}).

Overlooking issues of time in trading experiments is anachronistic. In 21st century financial markets, speed is king. 
Competition between automated trading systems (ATS), looking to capitalise on fleeting opportunities ahead of rivals, has resulted in a proliferation of high frequency trading (HFT) algorithms capable of executing many thousands of trades each second \cite{DuffinCartlidge18}.
Market dynamics reflect this general acceleration in trading speed: individual stocks frequently exhibit {\em ultra extreme events}, with ten percent price swings in less than one tenth of a second \cite{JohnsonEtal13}; {\em flash crashes} cause markets to lose a trillion dollars in value in five minutes \cite{BaxterCartlidge13}; and when an ATS {\em malfunctions}, it can lose hundreds of millions of dollars in under an hour and drive the owners into administration \cite{BaxterCartlidge13}.

Here, we address the gap between research and reality by exploring reaction speed and trading urgency on a suite of reference algorithms (AA, GDX, GVWY, SHVR, ZIC, and ZIP) using the {\em Bristol Stock Exchange} simulation platform (see \cite{Cliff18-BSE}).  
In Section~\ref{sec:background}, we review related work and introduce key economic concepts and technical details of the financial trading agents and simulation platform used in this study. 
Section~\ref{sec:reaction} explores {\em reaction time} of traders through a speed proportional selection mechanism. 
In Section~\ref{sec:urgency}, we explore {\em trading urgency}, and introduce a method for modelling trading urgency in ZIP traders, inspired by Gjerstad's ``pace'' parameter \cite{Gjerstad04}. 
Finally, in Section~\ref{sec:conclusion}, we discuss findings and conclude that since reaction speed and trading urgency can improve trading profits, and since time is such an important factor in real-world trading, the trading agents' research community will benefit from focusing more attention on these timing issues in future.

\ifnum\ANON=0 
    {This paper extends work by Hanifan and Cartlidge on {\em reaction time} 
    of traders, originally published in the 2020 Proceedings of the International 
    Conference on Agents and Artificial Intelligence (ICAART) \cite{HanifanCartlidge20}. Much of the background material presented in Section~\ref{sec:background} is reproduced from \cite{HanifanCartlidge20}, while Section~\ref{sec:reaction} presents a condensed version of the key results in \cite{HanifanCartlidge20}. Section~\ref{sec:urgency} presents entirely new experiments and results, exploring the consequences of adding ``urgency'' or ``pace'' to the ZIP strategy, and draws from Watson's MSc thesis \cite{Watson19}.}
    \else
    {This paper extends work by {\bf [ANON]} on {\em reaction time} 
    of traders, originally published in {\bf [CITE ANON]}.
    Much of the background material presented in Section~\ref{sec:background} is reproduced from {\bf [CITE ANON]}, while Section~\ref{sec:reaction} presents a condensed version of the key results in {\bf [CITE ANON]}. Section~\ref{sec:urgency} presents entirely new experiments and results on {\em trading urgency}.}
    \fi 

\section{\uppercase{Background}}
\label{sec:background}
\noindent 

\subsection{Trading Agent Experiments}
\label{sec:background-agents}
\noindent 
In 1962, Vernon Smith published his landmark study, which reported on a series of trading experiments conducted with small groups of untrained human participants (i.e., students) to investigate competitive market behaviours \cite{Smith62}. He was able to demonstrate that these simple simulations of financial markets produced surprisingly efficient equilibration behaviours, with trade prices quickly tending to the theoretical equilibrium value predicted by the underlying market supply and demand. Intriguingly, three decades later, Gode and Sunder were able to reproduce similar results, but this time using `zero intelligence' (ZI) algorithmic traders that generate random quote prices \cite{GodeSunder93}. Despite their simplicity, markets of ZIC traders (the letter `C' indicating traders are {\em constrained} to not make a loss) were shown to exhibit equilibration behaviours similar to that of humans, suggesting that intelligence is not necessary for competitive markets to behave efficiently: the market mechanism (the rules of the continuous double auction) performs much of the work. 

\begin{figure}[tb]
  \centering
   \includegraphics[width=0.5\linewidth]{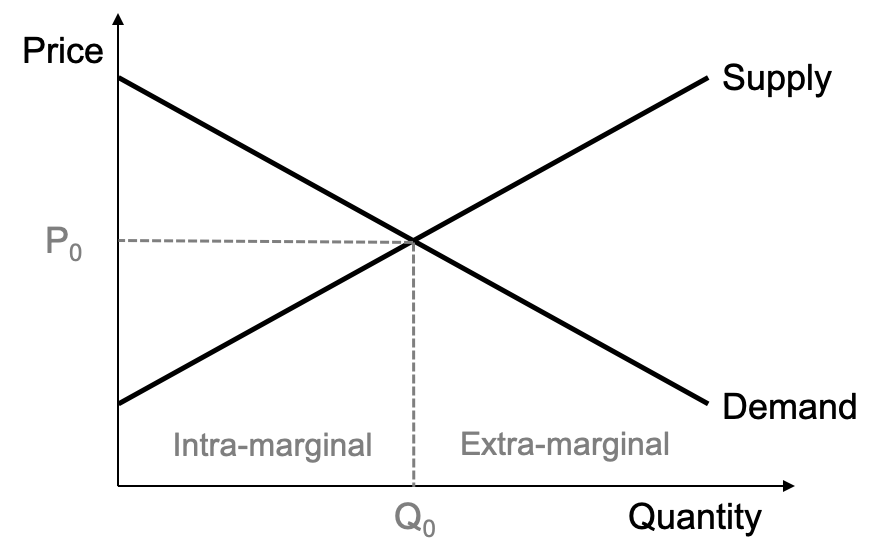}
  \caption{Supply and demand. The intersection of supply and demand determines the markets' competitive equilibrium price $P_0$ and quantity $Q_0$. {\em Intra-marginal} units to the left of $Q_0$ (i.e., quantity demanded with limit prices higher than $P_0$ and quantity supplied with limit prices lower than $P_0$) expect to transact; {\em extra-marginal} demand and supply will be unable to transact when the market is trading at equilibrium price.}
  \label{fig:DS}
\end{figure}

However, Gode and Sunder's result was later shown to only hold when market demand and supply are symmetric (i.e., when the magnitude of the gradient---the {\em price elasticity}---of supply and demand schedules are similar, such as the example shown in Figure~\ref{fig:DS}). For asymmetric markets, such as when the supply curve is horizontal, Cliff showed that `zero' intelligence is not enough to provide human-like levels of market efficiency \cite{Cliff97}. To account for this, Cliff introduced a new minimally-intelligent trading algorithm, which he named {\em Zero Intelligence Plus} (ZIP). ZIP maintains an internal profit margin, $\mu$, which is increased or decreased by traversing a decision tree that considers the most recent quote price, the direction of the quote (buy or sell) and whether it resulted in a trade. Margin, $\mu$, is then adjusted with magnitude proportional to a learning rate parameter, similar to that used in Widrow-Hoff or in back-propagation learning. Cliff successfully demonstrated that markets containing only ZIP traders will exhibit human-like behaviours in all of Smith's original experimental market configurations, both symmetric and asymmetric \cite{Cliff97}. 

\begin{figure}[tb]
  \centering
   \includegraphics[width=0.65\linewidth]{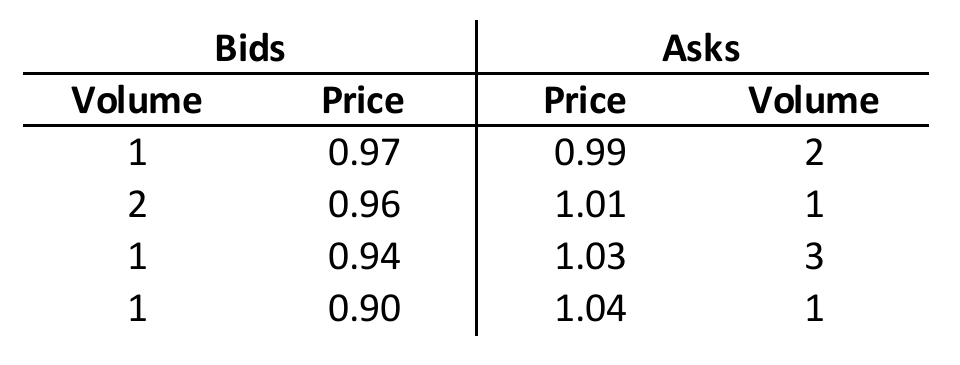}
  \caption{A Limit Order Book (LOB), presenting the current market state. Bids (orders to buy) are presented on the left hand side, ordered by price {\em descending}. Asks (orders to sell) are presented on the right hand side, ordered by price {\em ascending}.  Volume indicates the quantity available at each price. The top line presents the Best Bid ($BB=0.97$) and Best Ask ($BA=0.99$) prices in the market, and the difference between these prices is called the $spread=BA-BB=0.02$. The {\em midprice} of the book is $(BB+BA)/2=0.98$;
the {\em microprice} is volume weighted midprice, calculated as: $(2/3) 0.97+(1/3) 0.99=0.977$. Orders can be submitted at any price, subject to a minimum resolution, or tick size ($tick=0.01$). Aggressive orders that {\em cross the spread} (i.e., an ask with price $p_a\leq0.97$, or a bid with price $p_b\geq0.99$) will immediately execute at the price presented in the LOB (i.e., the ask will transact at price $BB=0.97$; the bid will transact at price $BA=0.99$). Passive orders that do not cross the spread will rest in the LOB, with position determined by price.  
    \ifnum\ANON=0 
    {(Reproduced from \cite{HanifanCartlidge20}.)}
    \else
    {(Reproduced from {\bf  [CITE ANON]}.)}
    \fi
  }
  \label{fig:LOB}
\end{figure}

Other intelligent trading agents have been developed to maximise profits in experimental markets that follow Smith's framework. Most notably, these include: GD, named after its inventors, Gjerstad and Dickhaut \cite{GD98}; and {\em Adaptive-Aggressive} (AA), developed by Vytelingum \cite{Vytelingum06}. GD selects a quote price by maximising a `belief' function of the likely profit for each possible quote, formed using historical quotes and transaction prices in the market. 
Over time, the original GD algorithm has been successively refined: first by Das et al. \cite{DasEtal01} and Tesauro and Das \cite{TesauroDas01} (named {\em Modified GD}, or MGD) to enable trading using an order book (see example in Figure~\ref{fig:LOB}), and to reduce belief function volatility; and then by Tesauro and Bredin \cite{TesauroBredin02}, who used dynamic programming to optimise cumulative long-term discounted profitability rather than immediate profit ({\em GD eXtended}, or GDX).  
In contrast, AA incorporates a combination of short-term and long-term learning to update an internal profit margin, $\mu$. In the short-term, $\mu$ is updated using rules similar to ZIP. Over the long-term, AA calculates a moving average of historical transaction prices to estimate the market equilibrium value, $P_0$, and current price volatility calculated as root mean square deviation of transaction prices around the estimate $P_0$. If the AA trader estimates that it is extra-marginal (and will therefore find it difficult to trade profitably: see Figure~\ref{fig:DS}) it trades more aggressively (by reducing $\mu$), if it is intra-marginal (and will therefore find it easier to profit) it trades more passively (by increasing $\mu$).  
For a summary of trading strategies, see Table~\ref{tab:traders}.

\begin{table*}[tb]
\vspace{3mm}
\caption{Summary of Trading Agent Strategies.}\label{tab:agents} \centering
{
\small
\setlength{\tabcolsep}{3pt} 
\renewcommand{\arraystretch}{1.15}
\begin{tabular}{lp{0.88\textwidth}}
  \toprule
  Trader & Method used by buyers to determine new bid price to quote, $Q$. \\
  \midrule
   AA & $Q=L(1-\mu)$, where  $0\leq\mu<1$ is an internal profit margin. Estimate $P_0$ to determine if $L$ is intra-marginal ($L\geq P_0$) or extra-marginal ($L<P_0$). If extra-marginal, increase aggressiveness (decrease $\mu$); else increase $\mu$. \cite{Vytelingum06}\\
   GDX & $Q$ selected by dynamic programming to maximise cumulative long-term discounted profitability of a belief function that calculates likely outcome of each price, $q$, based on the success of previous quotes and trade prices. \cite{TesauroBredin02}\\
  GVWY & $Q=L$. Always post quote at price equal to limit price, i.e., ``tell the truth''.\\
  SHVR & $Q=min(BB+T,L)$. Quote one tick inside current best bid. \\
  ZIC & $Q=q\in U[T,L]$. Quote selected randomly from a Uniform distribution with minimum value one tick ($T=0.01$) and maximum value limit price, $L$. \cite{GodeSunder93}\\
  ZIP & $Q=L(1-\mu)$, where  $0\leq\mu<1$ is an internal profit margin. 
  When a new trade occurs with price $p$, if $Q>p$ decrease $\mu$ (i.e., raise quote), else increase $\mu$. If new best bid has price $BB>Q$, decrease $\mu$ (i.e., raise quote). \cite{Cliff97}\\
  \midrule 
  \multicolumn{2}{p{0.98\textwidth}}{$Q$ is new quote price; $L$ is limit price; $T$ is tick size; $BB$ is best bid price on the LOB. Traders cannot make a loss, i.e., $Q\leq L$. Routines for sellers are symmetric to buyers.}\\ 
   \bottomrule
\end{tabular}
}
\label{tab:traders}
\end{table*}

\subsection{Trading Strategy ``Dominance''}\label{sec:background-dominance}
\noindent
For the last two decades, a research theme has emerged: to develop the best trading agent that can successfully beat human participants and other trading agents in Smith-style experiments (see \cite{SnashallCliff19} for a detailed historical account). In 2001, Das et al. first demonstrated that trading agents, specifically ZIP and MGD, outperform humans when directly competing in human-agent markets \cite{DasEtal01}.
This announcement quickly generated global media coverage and significant industry interest.
Shortly afterwards, Tesauro and Bredin \cite{TesauroBredin02} suggested that GDX {\em ``may offer the best performance of any published CDA bidding strategy''}. Subsequently, after its introduction in 2006 \cite{Vytelingum06}, AA was shown to dominate ZIP and GDX \cite{Vytelingum08} and also humans \cite{DeLucaCliff11}: {\em ``we therefore claim that AA may offer the best performance of any published strategy''}. And so, for several years, AA held the undisputed algo-trading crown.

\subsubsection{The King is Dead?}\label{sec:deadking}

More recently, the dominance of AA has been questioned in several works. Using the discrete-event simulation mode of OpEx, Vach \cite{Vach15} used simple Smith-style markets to compare efficiencies of traders in markets containing AA, GDX, and ZIP, as the proportion of each trader type in the market was varied. For large regions of this 3-trader mixture space, GDX was shown to be the dominant strategy, with AA only dominating in markets where there are significant proportions of other AA agents. This finding was supported by Cliff \cite{Cliff19}, through exhaustive testing of markets containing mixtures of MAA (a slightly {\em modified} version of AA which utilises {\em microprice} of the orderbook; see Figure~\ref{fig:LOB}), ZIC, ZIP, and SHVR (a simple non-adaptive strategy that quotes prices one tick inside the current best price on the order book; see Table~\ref{tab:traders}). 

Further, Cliff demonstrated that introducing more {\em realistic} market dynamics---continuous replenishment of assignments rather than periodic replenishments at regular intervals; and also a dynamic equilibrium, $P_0$, which was set to follow real world historical trade price data---MAA did {\em not} dominate, with SHVR and ZIP generating significantly more profits \cite{Cliff19}. 
A related study by Snashall and Cliff \cite{SnashallCliff19} also showed that GDX dominates in these more realistically complex markets containing ASAD ({\em Assignment-Adaptive}, developed by Stotter et al. \cite{Stotter13}), GDX, MAA, and ZIP traders; and moreover, GDX also outperforms AA in simpler Smith-style markets. 

In summary, these works suggest that AA's previously perceived dominance is sensitive to the mixture of competing strategies in the market, and the complexity of the underlying market dynamics. 

\subsection{Speed and Urgency}\label{sec:background-speed}
\noindent
Throughout the previous works, the primary motivation has been focused on pricing strategies for trading efficiency (i.e., profit maximisation and market equilibration behaviours). However, if we are to better understand the behaviour of these algorithms in more realistic environments, it is important to consider {\em latency}, a key real-world factor that is missing in most of these studies. In real-world financial markets, {\em communication latency} (the differential delays in which traders can access trading information and initiate trades with an exchange, e.g., \cite{MilesCliff19}), and {\em trading latency} (or {\em reaction time}: the time it takes for a human or algorithmic trader to react to new information) are major determinants of trading behaviours and market dynamics \cite{DuffinCartlidge18,SnashallCliff19}. In real markets, the proliferation and profitability of high frequency trading (HFT) evidences the efficacy of harnessing reduced latency, enabling traders to capitalise on fleeting opportunities ahead of competitors. McGroarty et al. introduced an agent model of financial markets with agents that operate on different timescales to simulate common strategies and behaviours, such as market makers, fundamental traders, high frequency momentum and mean-reversion traders, and noise traders \cite{McGroarty19}. 

\subsubsection{Reaction time:}
Several studies have conducted human-agent and agent-agent trading experiments using real-time asynchronous trading platforms. For their seminal demonstration of agents outperforming human traders, Das et al. \cite{DasEtal01} used a hybrid platform consisting of two of IBM's proprietary systems: GEM, a distributed experimental economics platform; and Magenta, an agent environment. Although real-time asynchronous, trading agents were constrained to operate on a sleep-wake cycle of $\bar{s}$ seconds, with {\em fast} agents having mean sleep time $\bar{s}=1$, and {\em slow} agents having mean sleep time $\bar{s}=5$. A random jitter was introduced for each sleep $s$ such that: $s\in[0.75\bar{s},1.25\bar{s}$]. Fast agents were set to wake on all new orders and trades, slow traders were set to wake only on trades. Therefore, although this real-time system enabled asynchronous actions, algorithmic traders were artificially slowed to have reaction times comparable with human traders. 

Following Das et al. \cite{DasEtal01}, other real-time human-agent experiments have invoked a similar sleep-wake cycle. Using the {\em Open Exchange} (OpEx) platform,\footnote{OpEx is available online: \url{https://sourceforge.net/projects/open-exchange}.} De Luca et al.  \cite{DeLuca11} demonstrated AA, GDX, and ZIP outperform humans when agents have sleep-wake cycle $\bar s=1$; agent-agent experiments, demonstrating AA dominance, were performed using a discrete event model (such that reaction times were ignored). OpEx has also been used for further human-agent experiments, for example, to demonstrate that: aggressive ({\em spread-jumping}) agents that are faster (i.e., those with lower $\bar{s}$ values) can perform less well against humans \cite{DeLuca11}; faster trading agents can reduce the efficiencies of human traders in the market \cite{Cartlidge12}; and agents with reaction speeds much quicker than humans can lead to endogenous fragmentation within a single market, such that fast (slow)
traders are more likely to execute with fast (slow) traders (\cite{CartlidgeCliff12}; a result that has analogies with the {\em robot phase transition} demonstrated in real-world markets \cite{JohnsonEtal13}). 

Agent-only real-time asynchronous experiments have also been conducted using the {\em Exchange Portal} (ExPo) platform.\footnote{ExPo is available online: \url{https://sourceforge.net/projects/exchangeportal}.} Stotter at al. \cite{Stotter13,Stotter14} used ExPo to introduce a new Assignment-Adaptive (ASAD) trading agent. They demonstrated that in ASAD:ZIP markets (with sleep-wake cycle, $\bar{s}=4$), signals produced by the trading behaviour of ASAD are beneficially utilised by ZIP traders, to the detriment of ASAD themselves. 

These works are representative of the literature relevant to reaction time and in automated trading algorithms. In general, we see that reaction times are either skewed by enforced sleep, directly encoded, or drawn from a probability distribution. As far as the authors are aware, there are no previously-published attempts to systematically understand the effects of reaction time using accurate computation times of individual trading strategies. 

\subsubsection{Urgency:} In March 1990, the Santa Fe Institute held a computerised discrete-time double auction tournament. Todd Kaplan, an economist at the University of Minnesota, won the competition with a simple rule-of-thumb strategy: {\em wait in the background and let others do the negotiating, but when bid and ask get sufficiently close, or when time is running out, jump in and `steal the deal'} \cite{rust94}. This strategy is now widely referred to as ``Kaplan's sniper''.
To steal the deal, Kaplan's snipers submit a bid (ask) with price equal to the last best ask (bid); thereby crossing the spread. Since Kaplan's snipers tend to wait in the background, this parasitic strategy does not help price formation.
Therefore, when too many traders follow the same strategy, market efficiency falls. However, in many markets, hiding in wait for an opportunity to present itself (i.e., {\em sniping}) can be profitable. 
Such behaviour is typically observed in online auction venues with fixed deadlines (such as eBay), where snipers wait until the auction is about to close before urgently posting bids. In contrast, a similar sniping effect is not observed in venues (such as Amazon) where auctions have no fixed deadline  \cite{OckenfelsRoth13}. 

Apart from Kaplan's sniper, the only other trading agent with an explicit sense of trading urgency is Gjerstad's extension of GD (named Heuristic Belief Learning, or HBL), which includes a ``pace'' parameter to alter the submission rate and quote price as a function of time remaining before deadline \cite{Gjerstad04}. In human-agent trials, Gjerstad demonstrated that the performance of faster paced HBL traders is comparable to that of humans; suggesting trading urgency can be beneficial \cite{Gjerstad04}.
Since urgency can be roughly approximated as aggressiveness (urgent traders anxious to trade will submit a more aggressive quote price), one could consider AA agents as having a sense of urgency, as aggressiveness is dynamically tuned based on prevailing market conditions \cite{Vytelingum06}. However, AA aggressiveness was not described as a mechanism for modelling temporal urgency, and does not explicitly consider trading deadlines.

In summary, we see that despite its importance in real world trading, the literature on urgency and reaction speed in financial trading agents is relatively sparse. In this paper, we attempt to address these gaps.

\section{\uppercase{Reaction Time}}
\label{sec:reaction}
\noindent 

\ifnum\ANON=0 
    {
    \noindent
    This section presents a condensed version of an exploration on {\em reaction time} of traders, originally published in the 2020 Proceedings of the International Conference on Agents and Artificial Intelligence (ICAART) \cite{HanifanCartlidge20}. For further exploration (including {\em fixed ordering} and {\em tournament selection} models), we refer the reader to \cite{HanifanCartlidge20} and also Hanifan's MSc thesis \cite{Hanifan19}.
    }
    \else
    {
    \noindent
    This section presents a condensed version of an exploration on {\em reaction time} of traders, originally published in {\bf [CITE ANON]}. For further exploration (including {\em fixed ordering} and {\em tournament selection} models), we refer the reader to {\bf [CITE ANON]} and also {\bf [CITE ANON]}.
    }
\fi 

\subsection{Modelling Reaction Time}
\label{sec:models}

For all experiments performed in this paper, we use the {\em Bristol Stock Exchange} (BSE), a teaching and research platform designed for running controlled financial trading experiments (for details, see \cite{Cliff18-BSE}). 
BSE is a minimal, discrete-time simulation of a centralised financial market, containing a single exchange with public limit order book (LOB). Trading experiments can be quickly configured by defining supply and demand schedules, the times that assignments are given to traders, and the strategies that each trader will follow. The BSE repository has reference implementations of a selection of trading strategies from the literature, including six that we consider in this paper: AA, GDX, GVWY, SHVR, ZIC, and ZIP. Python source-code for BSE is available open-source on the GitHub repository.\footnote{BSE is available online: \url{https://github.com/davecliff/BristolStockExchange}. In this paper, we use BSE version dated 22/07/18, with commit hash: \texttt{c0b6a1080b6f0804a373dbe430e34d062dc23ffb}.} 

\subsubsection{Random Order Selection:}
Each time step, BSE ensures that all traders have exactly one opportunity to act. This is achieved by selecting traders at random, and without replacement. 
When not selected, a trader cannot act. Therefore, when a profitable opportunity is presented in the market, a trader is likely to miss that opportunity unless it is lucky enough to be selected soon. 
Over a long simulation with many time steps, profitable opportunities will tend to be shared equally between traders, making random selection a fair process for comparing performance of strategies. However, in the real world, when a profitable opportunity is presented, traders that can recognise and act on that opportunity most quickly, are the most likely to profit. Therefore, the random order selection process is {\em unrealistic}, as it models all trading algorithms to have the same reaction speed. 

Random order selection suffers from the following unrealistic outcomes:
\begin{enumerate}
\item All traders have the same number of opportunities to act, irrespective of relative speed; and
\item Each trader has an equal chance to get {\em lucky} by being selected next. 
\end{enumerate}

\noindent 
To address these issues, we introduce a more {\em realistic} proportional selection model to simulate reaction time in the BSE framework.

\subsubsection{Speed Proportional Selection:}
\label{sec:method-time}
We assign each trader $t$ with a {\em reaction time} $R^t$ to represent the time taken for trader $t$ to react to new market information (i.e., $t$'s computation time). We adapt the BSE simulation so that, each time step, traders are selected to act in proportion to their relative speeds. For example, fast trader $F$ with reaction time $R^F=1$ acts twice as often as slow trader $S$ with reaction time $R^S=2$. To achieve this, we select traders from a biased pool containing multiple references to each trader, such that the number of references to trader $t$ is inversely proportional to $t$'s relative reaction time $R^t$. For example, if $R^F=1$ and $R^S=2$, we generate a biased pool, $P=\{F,F,S\}$, containing two references to fast trader $F$ and one reference to slow trader $S$. Each time step, traders are randomly selected, without replacement, until the pool is empty. 
We use notation $R^F_S=1/2$ to indicate $F$'s reaction time is half $S$'s reaction time (i.e., $F$ is {\em twice as fast} as $S$); similarly $R^S_F=2$ indicates $S$'s reaction time is twice as long as $F$'s (i.e., $S$ has {\em half the speed} of $F$). Therefore, for a market containing two strategies $F$ and $S$, with $R^F_S=1/n$, the biased pool $P$ will contain traders $F$ and $S$ in a ratio of $n:1$, respectively. 

Speed proportional selection provides several benefits of added realism:
\begin{enumerate}
\item faster agents have more opportunities to act than slower agents;
\item faster agents can act multiple times before a slower trading agent acts; and 
\item since ordering is random, slower traders can get {\em lucky} by being selected next, but the likelihood of this diminishes as the difference in relative speeds increase. 
\end{enumerate}
\noindent
Notice that, when all traders in market $M$ have the same relative speed (i.e., when $\forall i,j\in{M}, R^i_j=1$), the speed proportional selection model reduces to the BSE default random order selection model.

\subsection{Experimental Configuration}
\label{sec:config}

\begin{figure}[tb]
  \centering
   \includegraphics[width=0.65\linewidth]{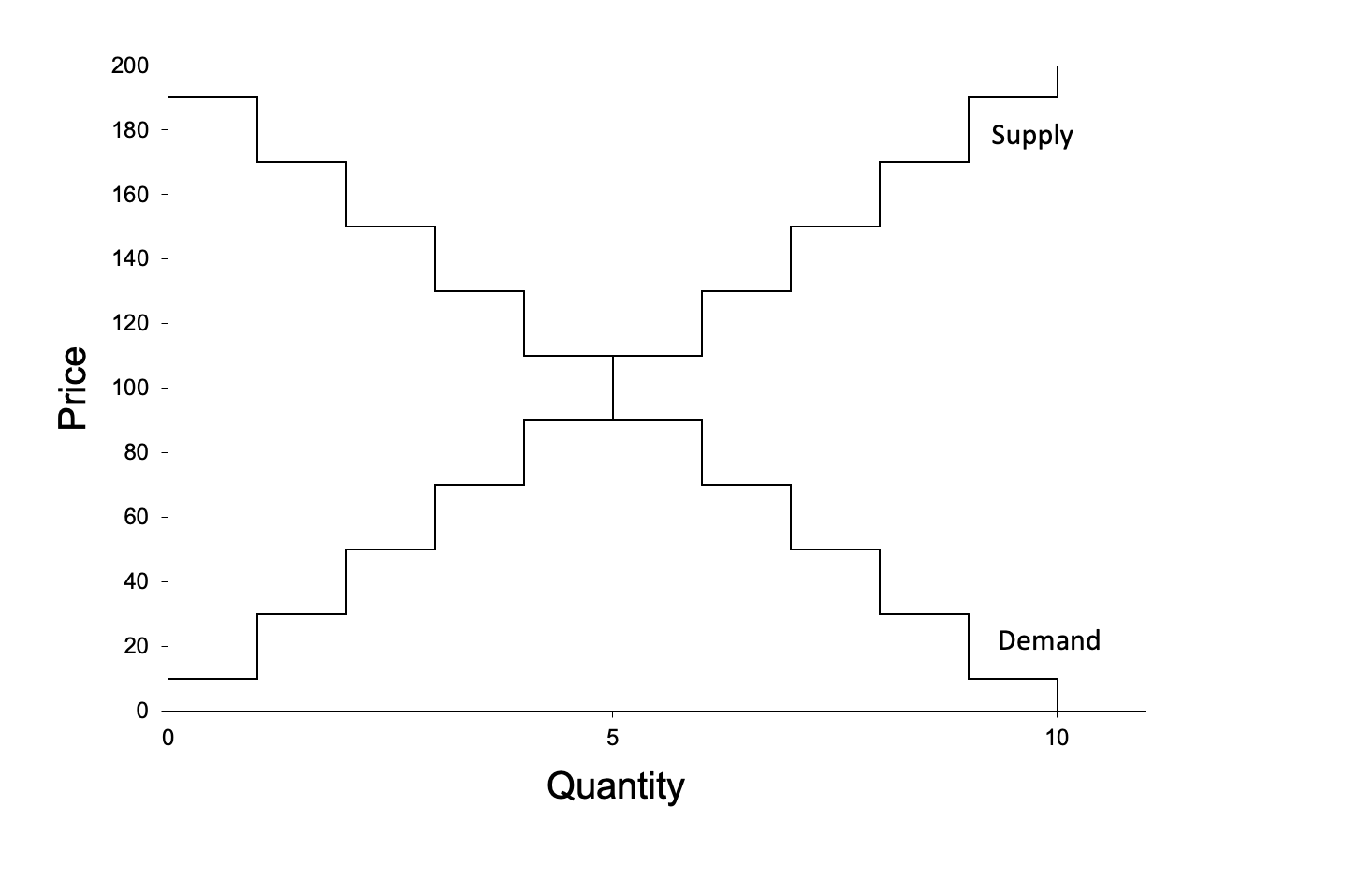}
  \caption{Symmetric supply and demand schedules for reaction speed experiments. The market has $n=10$ buyers (demand) and $n=10$ sellers (supply) with limit prices evenly distributed between 10 and 190, giving a theoretical equilibrium price $P_0=100\pm10$ and quantity transacted $Q_0=n/2=5$.
  }
  \label{fig:DS-reaction}
\end{figure}

\noindent
We performed a series of experiments, with markets containing an equal number of $n$ buyers and $n$ sellers, and with assignment limit prices distributed evenly between minimum value of 10 and maximum value of 190, as shown graphically in Figure~\ref{fig:DS-reaction}. Market sessions lasted 330 time steps, with assignments replenished periodically every 30 time steps.  This simple, static, symmetric market is deliberately chosen to enable comparisons with the literature. We use the speed proportional selection model (described in Section~\ref{sec:models}) to explore the effects of varying the reaction time of trading agents. Each simulation configuration was repeated 100 times, with results graphs plotting mean $\pm95\%$ confidence intervals. Where $p$ values are presented, statistical significance is calculated using Student's t-test. 

\subsection{Sensitivity Analysis}\label{sec:sensitivity}

\begin{figure*}[tb]
  \centering 
  \subcaptionbox{AA:GVWY}[0.45\linewidth][c]{%
    \includegraphics[width=\linewidth]{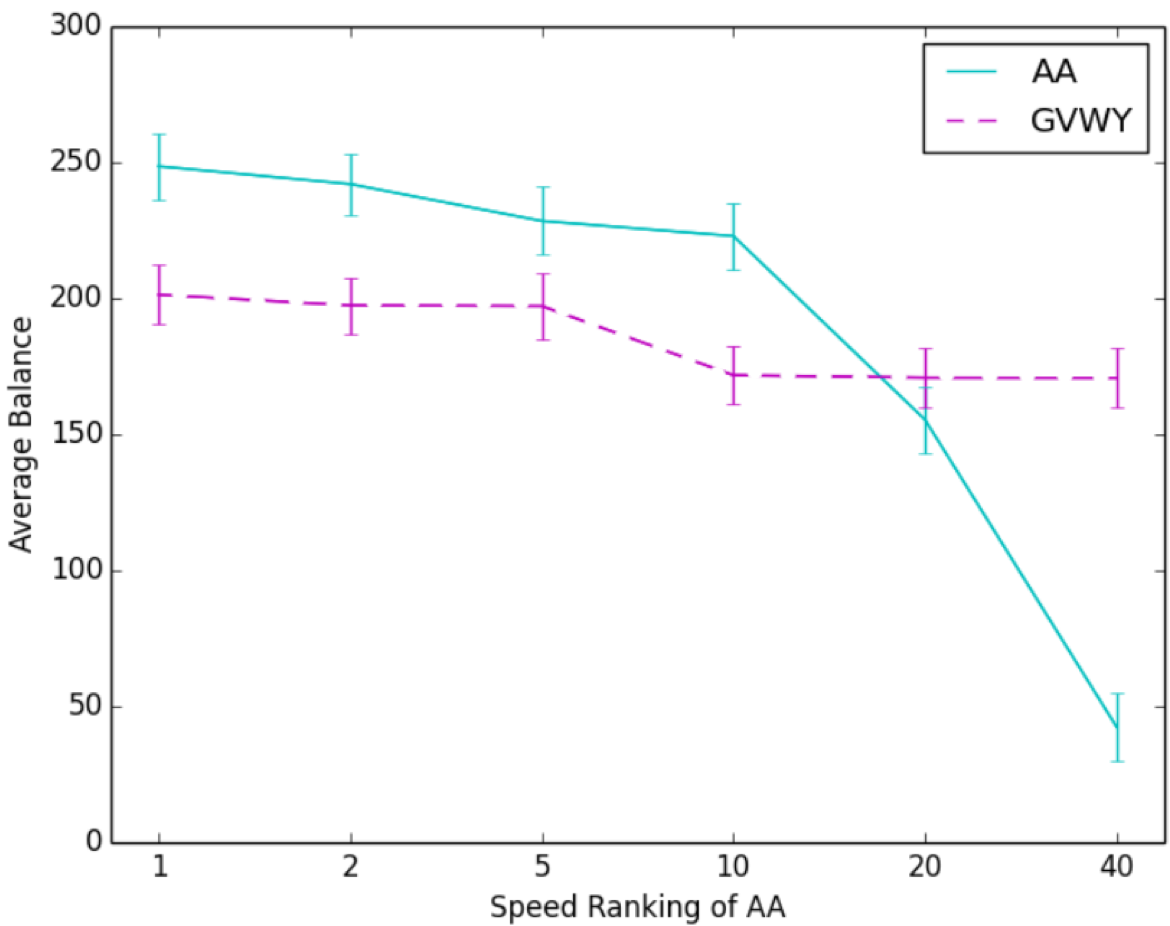}} 
    \hspace{0.05\textwidth}
  \subcaptionbox{AA:SHVR}[.45\linewidth][c]{%
    \includegraphics[width=\linewidth]{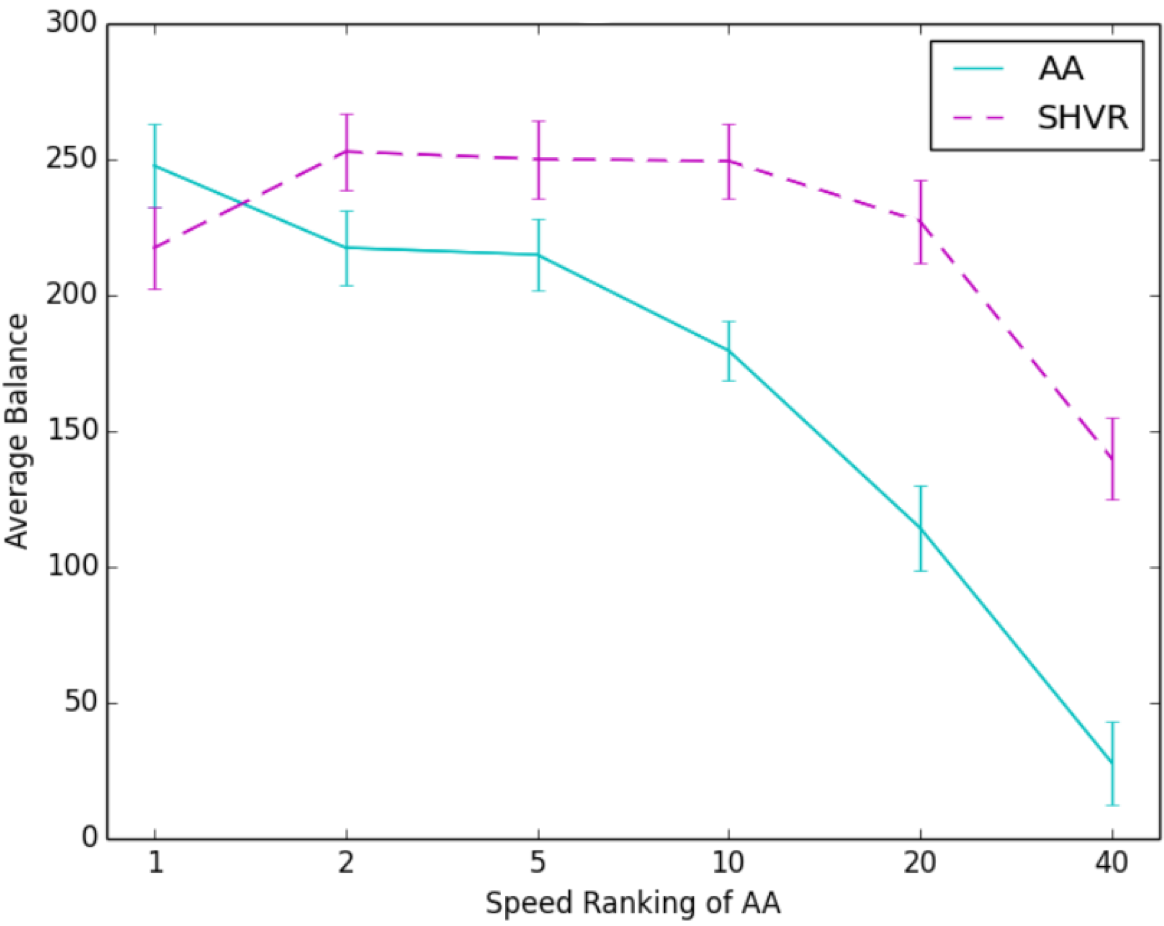}}
  \subcaptionbox{AA:ZIC}[.45\linewidth][c]{%
    \includegraphics[width=\linewidth]{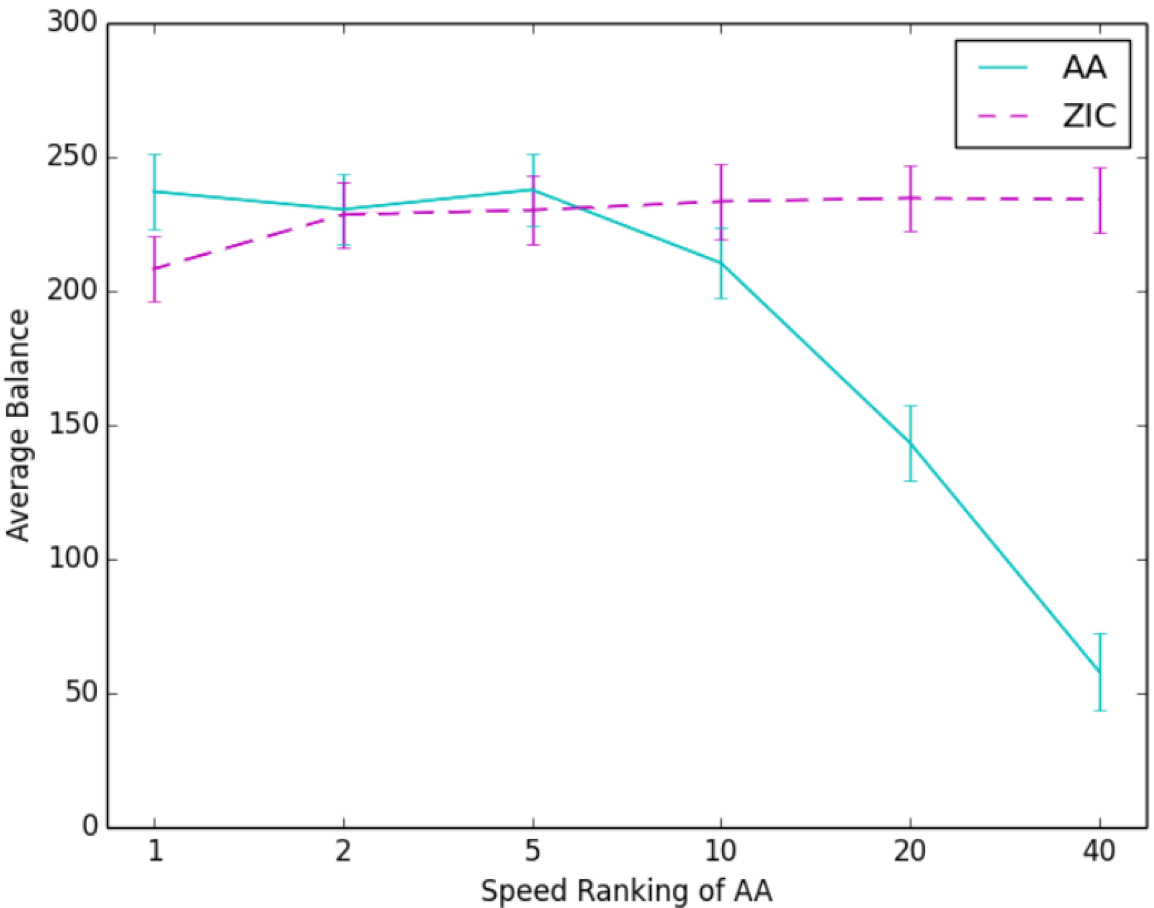}} 
    \hspace{0.05\textwidth}
  \subcaptionbox{AA:ZIP}[.45\linewidth][c]{%
    \includegraphics[width=\linewidth]{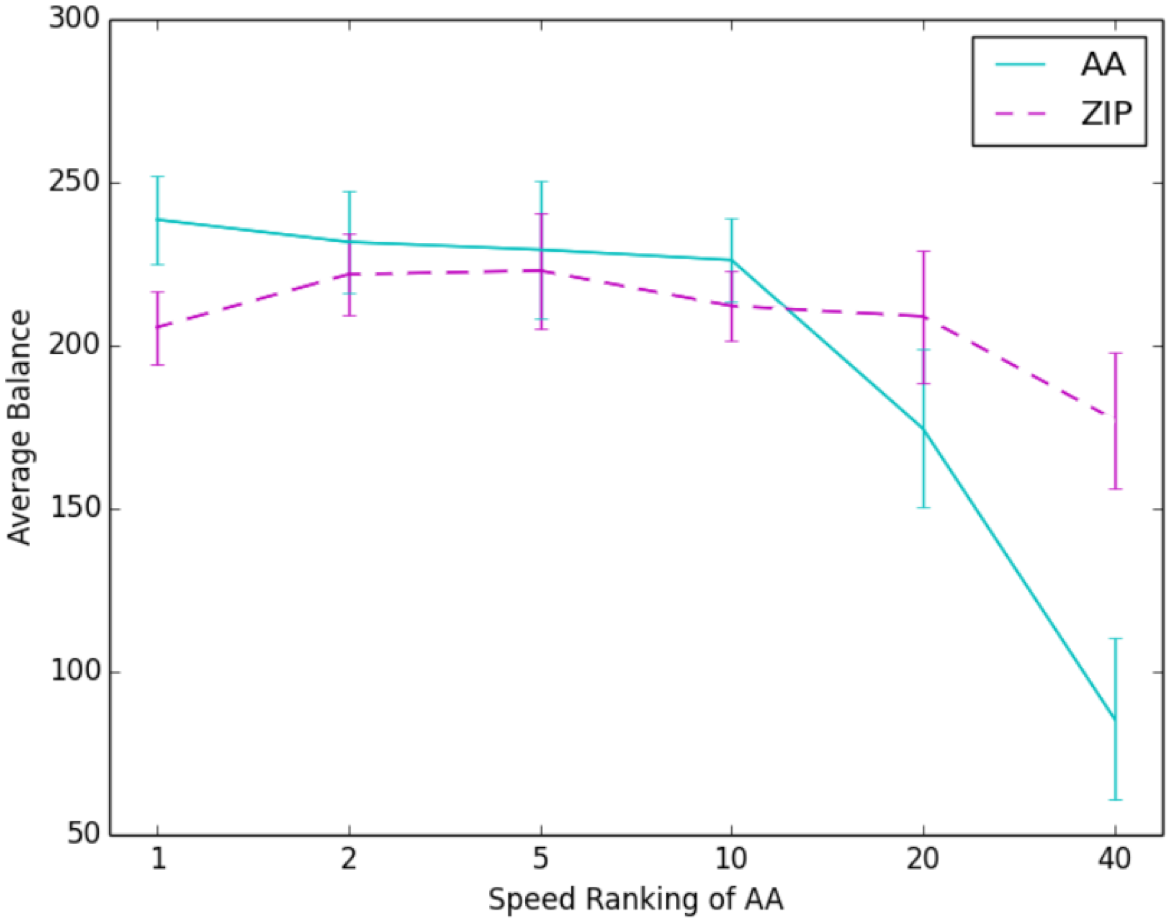}}
  \caption{Sensitivity analysis of AA using speed proportional selection in heterogeneous balanced tests. 
  Reaction time of AA relative to the competing trader type is varied from $R^{AA}_*=1$ to $R^{AA}_*=40$ (x-axis). 
  Each test, AA (light blue) outperforms the competitor (purple dash) when compute times are equal ($R^{AA}_*=1$). 
  As $R_*^{AA}$ increases, 
  AA performance falls, until an inversion point is reached where AA no longer outperforms the competitor. 
  For SHVR, inversion occurs between $1<R_{SHVR}^{AA}<2$. 
    \ifnum\ANON=0 
    {(Figures reproduced from \cite{HanifanCartlidge20}.)}
    \else
    {(Figures reproduced from {\bf  [CITE ANON]}.)}
    \fi
  }
  \label{fig:AA-timing}
\end{figure*}

Figure~\ref{fig:AA-timing} presents results of speed sensitivity analysis of AA performance in heterogeneous balanced-group tests against each of four other trading strategies: (a) GVWY, (b) SHVR, (c) ZIC, (d) ZIP. The reaction time of AA relative to the competing trader strategy is varied from $R^{AA}_*=1$ (i.e., equal speed) to $R^{AA}_*=40$ (i.e., AA is forty times {\em slower} than the competing strategy). Graphs show the effect of increasing $R^{AA}_*$ (x-axis: as we move right, AA is increasingly slowed). In each case, we see that when $R^{AA}_*=1$ (i.e., equal reaction times, equivalent to the default random order selection model used in the literature), AA (light blue line) outperforms the competing trader (purple dashed line). This is the BSE default setting, and the result confirms previous findings, which suggest $AA$ dominates in symmetric markets with balanced numbers of traders \cite{DeLucaCliff11}.

However, as relative AA reaction time $R^{AA}_*$ is increased (i.e., as AA becomes relatively {\em slower}), we see that AA performance gradually falls, until a point is reached where $AA$ performs {\em worse} than the strategy it is competing against. This inversion point varies depending on which pair of strategies are being tested: (a) for GVWY, inversion is between $10<R^{AA}_{GVWY}<20$; (b) for SHVR, between $1<R_{SHVR}^{AA}<2$; (c) for ZIC, between $5<R^{AA}_{ZIC}<10$; and (d) for ZIP, between $10<R^{AA}_{ZIP}<20$. 

Most noticeable is the low inversion value for SHVR (between 1 and 2), 
suggesting that AA's dominance over SHVR is particularly sensitive to small variations in relative trader speeds.

\begin{table}[tb]
\vspace{2.5mm}
\caption{Profiled reaction times
    \ifnum\ANON=0 
    {(reproduced from \cite{HanifanCartlidge20}).}
    \else
    {(reproduced from {\bf  [CITE ANON]}).}
    \fi
  }
  \label{tab:profile} 
  \small
  \centering
  \setlength{\tabcolsep}{3pt} 
  \renewcommand{\arraystretch}{1.15}
  \begin{tabular}{ccccc}
  \toprule
  Trader  & Time ($\mu$s) & Stateful & Reactive & $R^*_{SHVR}$\\
  \midrule
  GVWY & 4.2 & N & N & 0.61\\
  SHVR  & 6.9 & N & Y & 1.00\\ 
  ZIC      & 7.1 & N & N & 1.03\\
  ZIP      & 8.4 & Y & Y & 1.22\\
  AA       & 9.5 & Y & Y & 1.38\\
  \bottomrule
\end{tabular}
\end{table}

\subsection{Profiling Reaction Times}
\label{sec:profiling} 

Having demonstrated that trading performance is sensitive to relative speed, we attempt to accurately profile the reaction times of each trading agent strategy. In BSE, the computation time of an agent is composed of two methods: 

\begin{description}
\item [\texttt{getOrder}] called when a trader is selected to act. This requires the calculation of a new quote price, $Q$, if a new order is to be submitted into the market;
\item [\texttt{respond}] called after each market event (e.g., a new trade, or a new best bid or ask on the LOB). Traders can use the event data to update any internal parameters they have (such as a profit margin). 
\end{description}

GVWY, SHVR, and ZIC are {\em stateless} traders: strategies have no internal parameters and therefore no action is taken when the \texttt{respond} method is called. In comparison, ZIP and AA are {\em stateful}: strategies use the \texttt{respond} method to update internal variables and to calculate a new profit margin, $\mu$. When \texttt{getOrder} is called, ZIP and AA use their current profit margin, $\mu$, to calculate a new quote price; SHVR uses the current best bid (or best ask) in the LOB to generate a new quote price; while the {\em nonreactive} strategies, GVWY and ZIC, generate a quote price without making reference to market data.  

We profiled the reaction times of each trading agent across a variety of market conditions, including varied population size, mix of trading strategies in the market, assignment replenishment schedules, supply and demand schedules, etc. 
\ifnum\ANON=0 
(see \cite{Hanifan19} for details). 
\else
(see {\em redacted} for details).
\fi
In Table~\ref{tab:profile}, we present mean time calculated across more than 52 million method calls. 
Unsurprisingly, we see that the traders with the longest computation times, ZIP and AA, are those with an internal state that requires continuous updating. 
The relative reaction times between the fastest and slowest traders is roughly a factor of two: $R_{GWVY}^{AA}=2.26$. 
The relative reaction times between AA and ZIP is $R^{AA}_{ZIP}=1.13$. This result is consistent with the average relative time $R^{AA}_{ZIP}=1.19$ presented by \cite{SnashallCliff19}, and we take this as confirmatory evidence that our profiling is accurate. The final column of Table~\ref{tab:profile}, headed $R^*_{SHVR}$, presents the reaction time of each trader relative to SHVR. Generating a quote price relative to the current LOB, SHVR is the only stateless (and therefore {\em fast}) trader that {\em reacts} to market information; although it does so in a simplistic non-adaptive fashion (unlike the slower AA and ZIP).

\begin{figure*}[tb]
  \centering 
  \label{subfig:a}
  \subcaptionbox{ZIP:SHVR\label{fig:hetero-time-a}}[0.45\linewidth][c]{%
    \includegraphics[width=\linewidth]{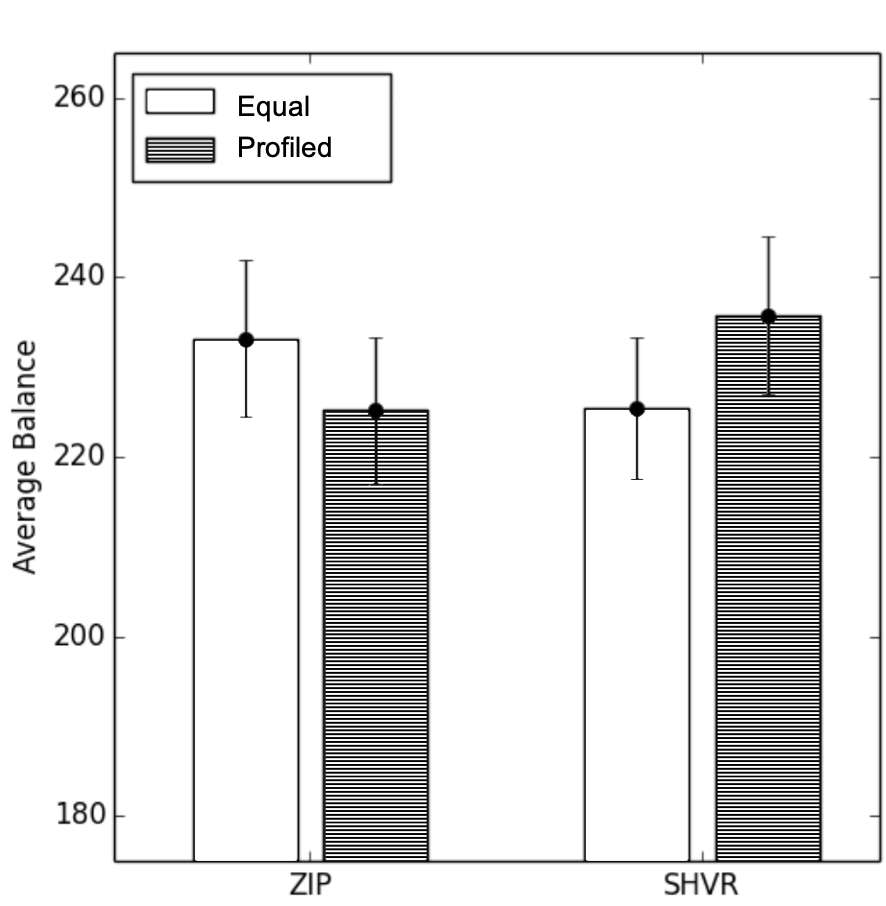}} 
    \hspace{0.03\textwidth}
  \subcaptionbox{AA:SHVR\label{fig:hetero-time-b}}[0.45\linewidth][c]{%
    \includegraphics[width=\linewidth]{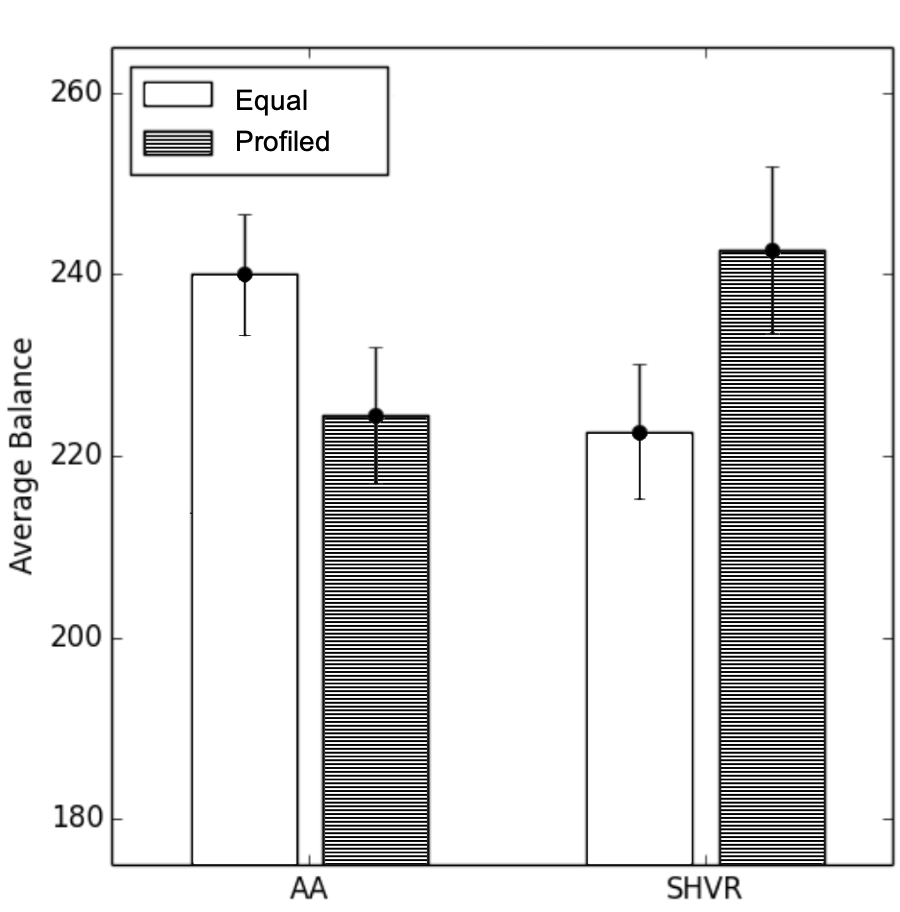}}
  \par\smallskip
  \caption{When traders have equal speed (white bars), SHVR is outperformed by ZIP (not significant) and AA (significant). When traders have profiled speeds (grey bars), SHVR outperforms ZIP (not significant) and AA (significant). }
  \label{fig:hetero-time}
\end{figure*}

\subsection{Results Using Profiled Reaction Times}
We used profiled computation times (see Table~\ref{tab:profile}) for proportional selection in heterogeneous balanced-group tests for pairwise comparisons between all trader types. The majority of results showed no significant difference, suggesting the relative differences in reaction speeds between the trader agents are not large enough to have an impact. However, results for ZIP:SHVR and AA:SHVR were particularly interesting (see Figure~\ref{fig:hetero-time}). 
For ZIP:SHVR (Figure~\ref{fig:hetero-time-a}), under BSE's default random order selection process (white bars), ZIP outperforms SHVR. However, when selecting traders proportional to their true relative speeds (grey bars) SHVR outperforms ZIP (although the difference is not significant; $p>0.05$). A similar, but more pronounced trend emerges between AA:SHVR (Figure~\ref{fig:hetero-time-b}). Here, AA significantly ($p<0.05$) outperforms SHVR under the default randomised selection (white), and significantly ($p<0.05$) underperforms SHVR under speed proportional selection (grey). 

This is a novel result. 
By accurately accounting for the relative reaction times of the two algorithms, we have demonstrated that, in balanced tests, 
SHVR---the simple non-adaptive order book strategy---is able to generate more profit than AA in a Smith-style static symmetric marketplace; exactly the kind of market that AA was specifically designed to succeed in \cite{Vytelingum06}, and in which several studies have previously demonstrated AA as being the dominant known strategy \cite{DeLucaCliff11,Vytelingum08}. 

We believe that this finding is significant, not only because it contributes to the recent body of evidence suggesting that AA is non-dominant (refer to Section~\ref{sec:deadking}), but also because it demonstrates that the performance of adaptive trading algorithms (AA and ZIP) are sensitive to reaction time; and once reaction time is considered, SHVR may be relatively superior. In more complex markets designed to emulate real-world financial dynamics, SHVR has previously been shown to outperform AA and to perform similarly to ZIP \cite{Cliff19}. Here, we extend this result to show that SHVR can also outperform in simple markets, once we account for speed. 

In this study, we have not considered the GDX trading strategy and we reserve this for future work. However, we note that Snashall and Cliff \cite{SnashallCliff19} have recently published profiled reaction times of GDX and report it to be an order of magnitude slower than AA (i.e., $R^{GDX}_{AA}\approx10$), due to its relatively complex internal optimisation process. 
It is therefore likely that GDX would also perform less well than SHVR (and AA and ZIP), once we accurately account for reaction times, as each competing strategy would have roughly ten opportunities to act, for every GDX trader's action. 


\section{\uppercase{Trading Urgency}}\label{sec:urgency}

Here, we turn our attention to {\em trading urgency} and return to the default random order selection method of BSE. The speed proportional selection model introduced in Section~\ref{sec:reaction} is not used here. 
\ifnum\ANON=0 
    {For further details on experiments and results presented in this section, we refer the reader to Watson's MSc thesis \cite{Watson19}.}
    \else
    {For further details on experiments and results presented in this section, we refer the reader to {\bf [CITE ANON]}.}
\fi 

\subsection{Pace}

To introduce trading urgency, we follow Gjerstad's ``pace'' approach for modelling wait times between posting orders in the HBL model \cite[p.14]{Gjerstad04}. Assume $\kappa-1$ asks/bids have occurred at times $\{t_1, t_2, \ldots t_{\kappa-1}\}$. Each trader $i$ calculates a time $t_{\kappa_i}$ to wait until posting next bid/ask, described by the following probability distribution: 

\begin{equation}
p_r [t_{\kappa_i} < t_{\kappa_{i-1}} + \tau ] = 1 - e^{-\tau/\lambda_i} 
\label{eq:prob}
\end{equation}

\noindent
Parameter $\lambda_i$ depends on trader $i$'s current maximum expected surplus $S_i^*$; current time that has elapsed in the trading period $t_{k-1}$; and on total time in the trading period $T$. Specifically:
\begin{equation}
\lambda_i(S_i, t_{\kappa-1},T) = \frac{ \beta_i(T-\alpha_i t_{\kappa-1})}{S_i^*T}
\label{eq:lambda}
\end{equation}
\noindent
where $\alpha_i\in(0,1)$ determines trader $i$'s acceleration in pace as the trading period progresses; $\beta_i\in(0,\infty)$ is a linear scale factor for the timing decision (where $\beta_i=250$ is ``fast'', and $\beta_i=400$ is ``slow''); and $S_i^*$ is the maximum expected surplus for agent $i$. Therefore, $\lambda_i$ signifies the mean wait time between $i$'s orders, and decreases in $S_i^*$, as traders with greater expected surplus are more anxious (i.e., have more {\em urgency}) to trade.

\subsection{ZIP with Pace (ZIPP)}

We modify ZIP to have urgency by incorporating Gjerstad's {\em pace} formulation. However, since ZIP does not calculate a maximum expected surplus, we instead calculate $S_i^*$ as trader $i$'s current surplus (i.e., profit). We name this algorithm ZIP-Pace, or ZIPP, and use parameter values $\alpha=0.95$ and $\beta=400$, which Gjerstad demonstrated to yield highest profits \cite{Gjerstad04}.

Each time ZIPP trader $i$ submits a new order, a maximum wait time $t_{\kappa_i}$ is generated using equations (\ref{eq:prob}) and (\ref{eq:lambda}). Trader $i$ must submit a new order {\em before} time $t_{\kappa_i}$ is reached. Either: (i) market updates trigger a new order submission from trader $i$ before $t_{\kappa_i}$ (using the standard ZIP logic), in which case a new maximum wait time $t_{\kappa_i}$ is generated; or (ii) if no order is submitted before $t_{\kappa_i}$, then at time $t_{\kappa_i}$ trader $i$ is forced to submit a new ``urgent'' order. This stops ZIPP traders from waiting for long periods between posting orders when there is little or no market activity, and enables ZIPP traders to react with urgency as the market close (or some other client-imposed) deadline approaches. 

When ZIPP trader $i$ posts an {\em urgent} order (i.e., when time $t_{\kappa_i}$ is reached), a new (more {\em aggressive}) quote price $q_i^{t}$ is calculated by shaving $\delta_i$ off the previous quote price $q_i^{t-1}$. For sellers, the quote price is decreased, i.e., $q_i^{t}=q_i^{t-1}-\delta_i$; for buyers, the quote price is increased, i.e., $q_i^{t}=q_i^{t-1}+\delta_i$. For trader $i$, the amount to shave $\delta_i$ is defined as: 

\begin{equation}
\delta_i = \frac{|L_i-q_i^{t-1}|}{T-t}
\end{equation}

\noindent
where $L_i$ is current limit price, $q_i^{t-1}$ is the previous quote price, $T$ is auction close (or {\em deadline}), and $t$ is current time. Therefore, $\delta_i$ increases as $t$ approaches $T$ (i.e., as auction close draws near), and decreases as $q_i^{t-1}$ approaches $L_i$ (i.e., as quote price nears limit price). Since minimum tick size in BSE is 1, if $\delta_i<1$, we set $\delta_i=1$; and to ensure traders do not make a loss, $q_i^{t}$ is restricted to a maximum bid value (or minimum ask value) equal to $L_i$.

\begin{figure*}[tb]
  \centering 
  \subcaptionbox{ZIP
  \label{fig:bid-prices-a}}[0.48\linewidth][c]{%
    \includegraphics[width=\linewidth]{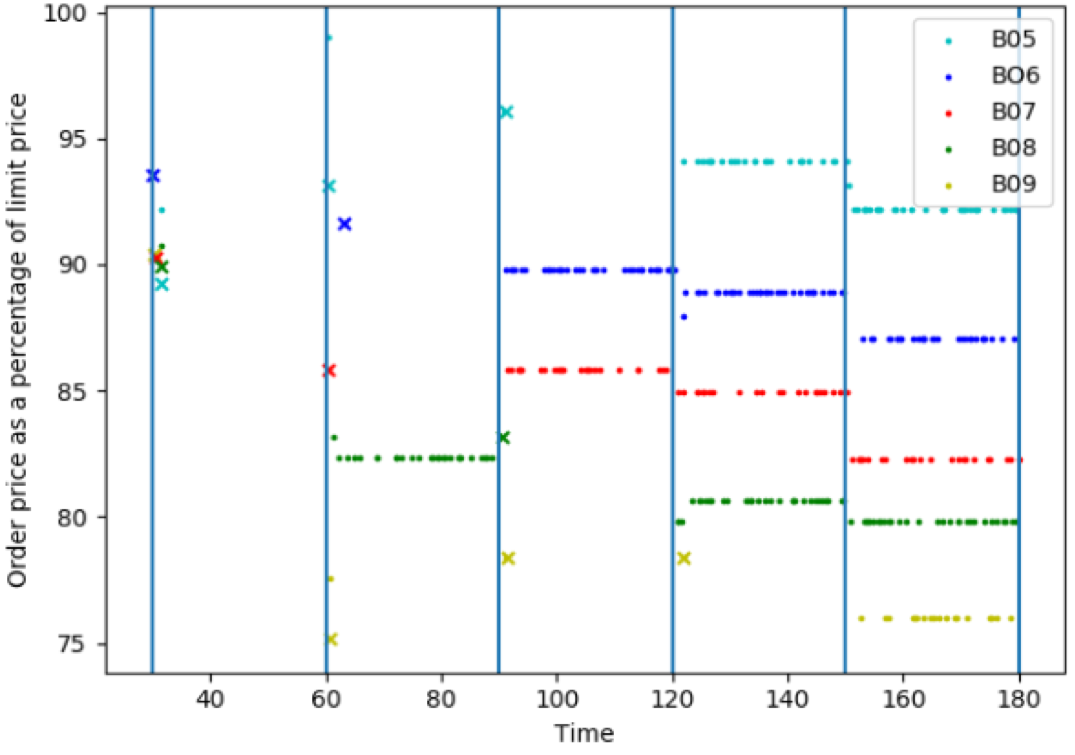}} 
    \hspace{0.01\textwidth}
  \subcaptionbox{ZIPP
  \label{fig:bid-prices-b}}[0.48\linewidth][c]{%
    \includegraphics[width=\linewidth]{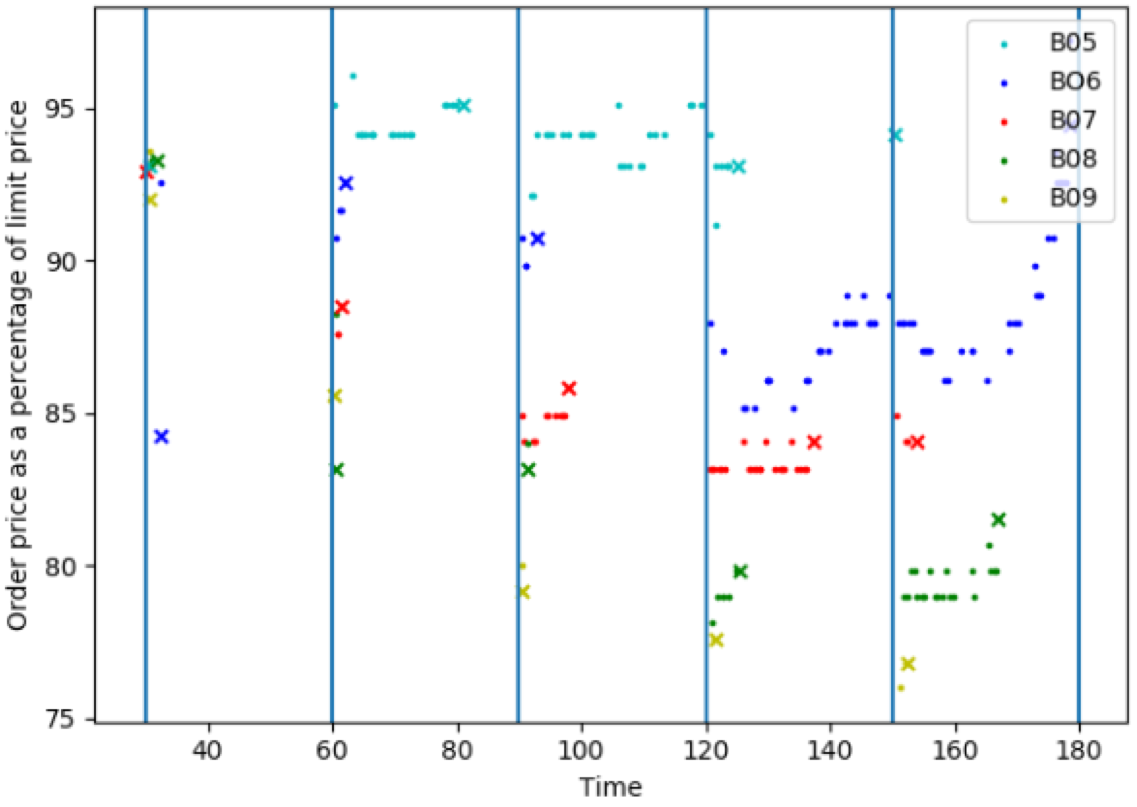}}
  \caption{Bid quote prices, $q_i$, as a percentage of limit price, $L_i$; i.e., $y=100 q_i/L_i$.}
  \label{fig:bid-prices}
\end{figure*}

\subsection{ZIPP Bidding Behaviour}

\begin{figure*}[tb]
  \centering 
  \subcaptionbox{Symmetric Market
  \label{fig:ben-sd-a}}[.3\linewidth][c]{%
    \includegraphics[width=\linewidth]{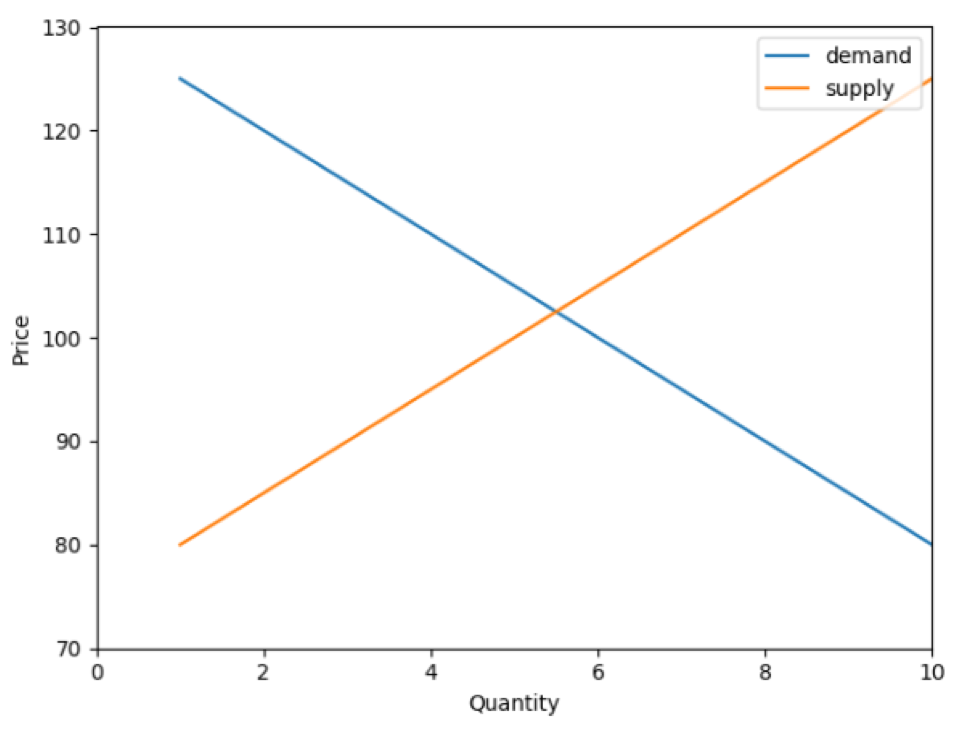}} 
    \hspace{0.02\textwidth}
  \subcaptionbox{Elastic Demand
  \label{fig:ben-sd-b}}[.3\linewidth][c]{%
    \includegraphics[width=\linewidth]{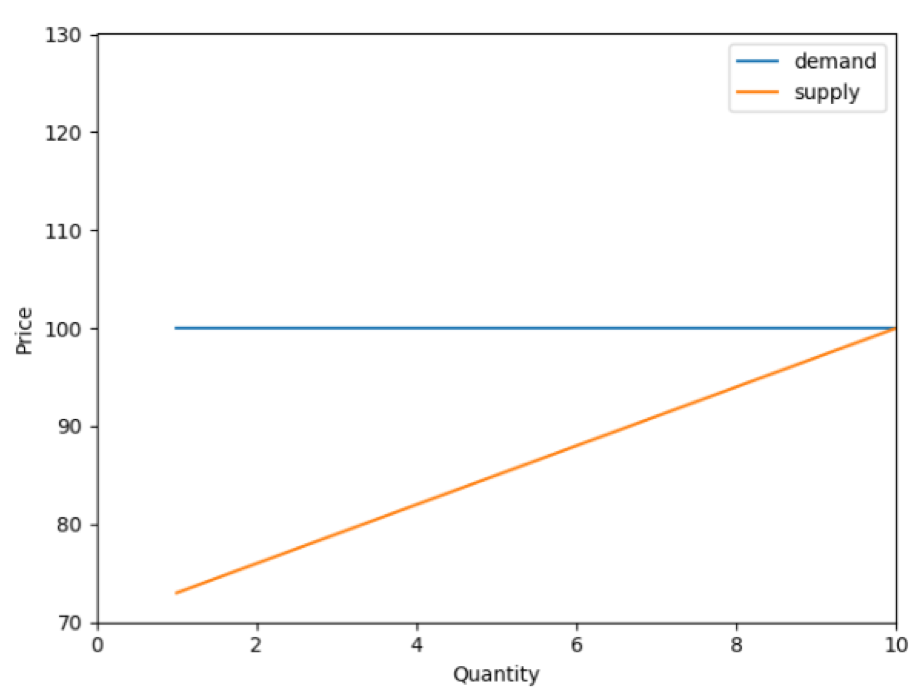}}
    \hspace{0.02\textwidth}
  \subcaptionbox{Elastic Supply
  \label{fig:ben-sd-c}}[.3\linewidth][c]{%
    \includegraphics[width=\linewidth]{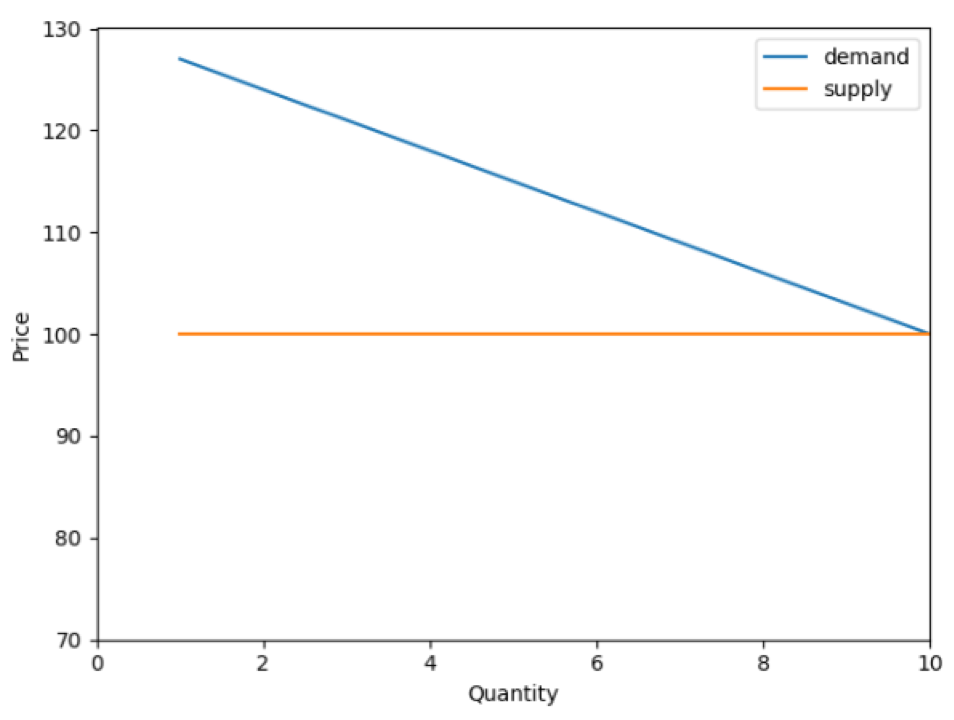}} 
  \subcaptionbox{Excess Demand
  \label{fig:ben-sd-d}}[.3\linewidth][c]{%
    \includegraphics[width=\linewidth]{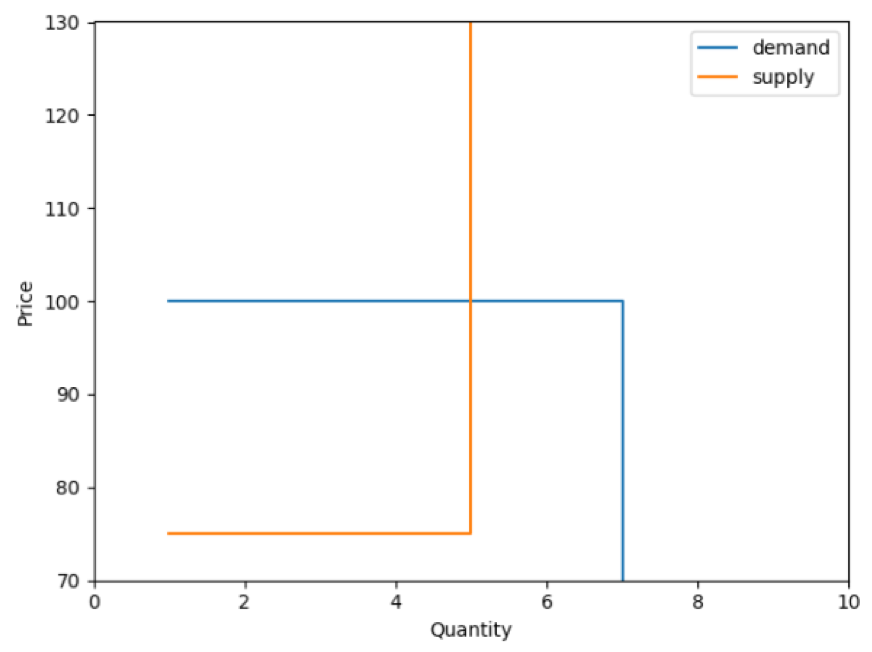}}
    \hspace{0.02\textwidth}
  \subcaptionbox{Excess Supply
  \label{fig:ben-sd-e}}[.3\linewidth][c]{%
    \includegraphics[width=\linewidth]{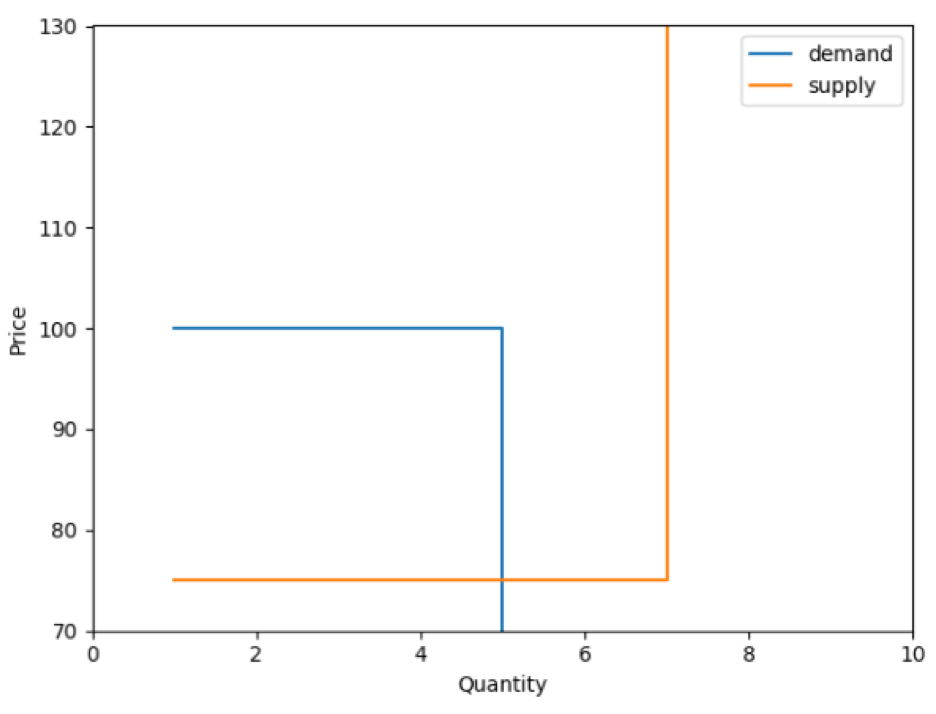}}

  \caption{Supply and demand schedules used for trading urgency experiments.}
  \label{fig:ben-sd}
\end{figure*}

To demonstrate the behaviour of ZIPP, we compare homogeneous markets: (i) containing only ZIP traders (Figure~\ref{fig:bid-prices-a}); and (ii) containing only ZIPP traders (Figure~\ref{fig:bid-prices-b}).
Each market contains 20 traders (10 buyers and 10 sellers), with assignments distributed using symmetric demand and supply schedules (see Figure~\ref{fig:ben-sd-a}) with minimum / maximum limit order values $(L_{min},L_{max})=(75,125)$. 
A trading day lasts 180 seconds, and each trader receives a new assignment to trade every 30 seconds (we call this the {\em assignment period}, represented by blue vertical lines). 
Therefore, traders receive 5 assignments to trade each day, and if an assignment has not been completed before a new assignment is received (i.e., before the assignment period ends), the original assignment is cancelled and opportunities for making profit from the assignment are lost. 
Figure~\ref{fig:bid-prices} presents the trading behaviours of the 5 intra-marginal buyers (i.e., those {\em expected} to trade) in each market (note that behaviours for sellers, not shown, are symmetric). The y-axis plots trader $i$'s quote price $q_i$ as a percentage of limit price $L_i$, such that $y=100(q_i/L_i)$, with a dot representing an unexecuted order (resting in the order book), and a cross representing order execution. 
In each assignment period, there can be at most one order execution per trader. 

In Figure~\ref{fig:bid-prices-a}, ZIP traders begin by posting quotes with relatively small margin (around 90\% of limit price, or above), which quickly results in all traders executing their orders (crosses). In each subsequent assignment period, ZIP traders tend to increase their margin (and therefore post a lower quote price), with little or no variation in quote price during the period (indicated by horizontal banding of dots). As a result, during the final period, no ZIP trader executes a trade, even though each has a significant profit margin that can be relaxed (B09 continues to maintain a margin of nearly 25\% throughout the final period). In contrast, ZIPP traders (Figure~\ref{fig:bid-prices-b}) continue to explore different quote prices during each assignment period (indicated by horizontal zig-zagging of dots), with every trader executing a trade (cross) in the final period. Notice how B06 increasingly reduces margin during the final period (from 13\% to 5\%) as the deadline approaches, eventually executing a trade shortly before close (cross obscured by legend). Overall, homogeneous markets of ZIPP traders execute more transactions than homogeneous markets of ZIP traders, resulting in a more efficient market and greater profit for ZIPP traders. This is the behaviour we desire.

\subsection{ZIPP Performance Testing}

\subsubsection{Method:}

For testing the performance of ZIPP (and comparing performance against other trading agents), we use five different supply and demand schedules (see Figure~\ref{fig:ben-sd}), each displaying different market properties: (a) symmetric supply and demand; (b) price elastic demand, with all buyers having the same limit price; (c) price elastic supply; (d) an excess demand, with more buyers than sellers; and (e) excess supply, with more sellers than buyers.  Each market contains exactly 10 buyers and 10 sellers. In homogeneous markets we have 20 traders of one type. In heterogeneous balanced-test markets, containing multiple trader types, we have 5 buyers of each type, and 5 sellers of each type. Each test is run over one day, lasting 180 seconds, with assignments distributed every 30 seconds (i.e., each day contains five assignment periods). For each condition, we perform 25 repeated trials. 

\subsubsection{Homogeneous Markets:}

\begin{figure*}[tb]
  \centering 
  \subcaptionbox{ZIP vs. ZIPP
  \label{fig:ben-results-homo-a}}[0.48\linewidth][c]{%
    \includegraphics[width=\linewidth]{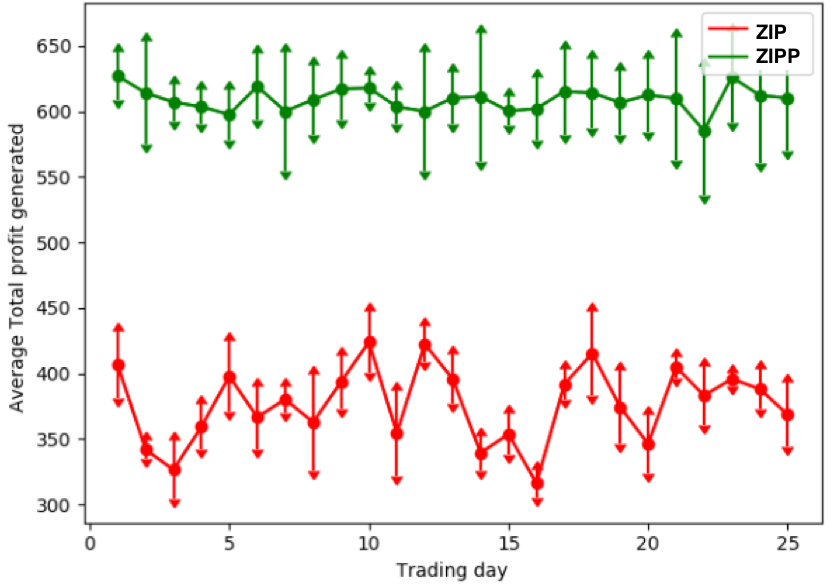}} 
    \hspace{0.01\textwidth}
  \subcaptionbox{ZIP, ZIPP, AA, GDX
  \label{fig:ben-results-homo-b}}[0.48\linewidth][c]{%
    \includegraphics[width=\linewidth]{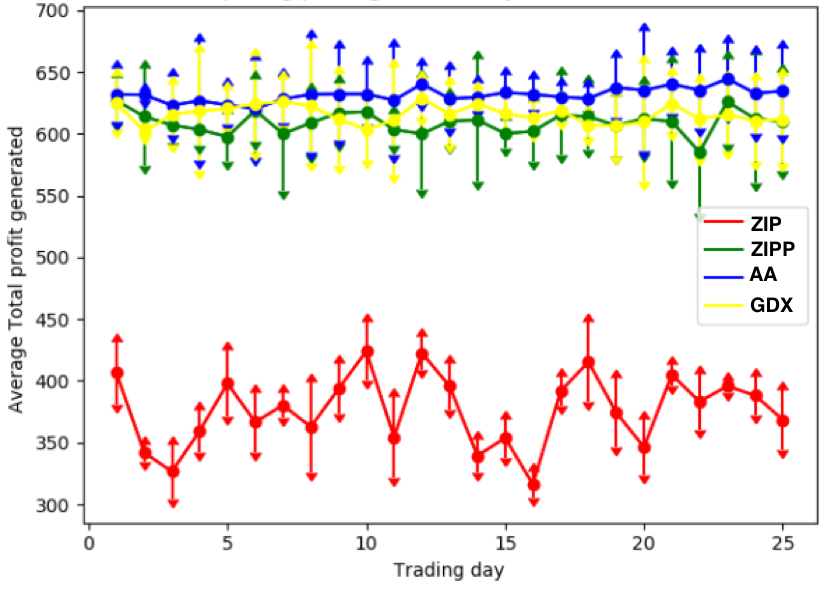}}
  \caption{Average total profit generated each trading day (homogeneous markets).}
  \label{fig:ben-results-homo}
\end{figure*}

Figure~\ref{fig:ben-results-homo} shows mean profits ($\pm$ standard deviation) generated in homogeneous markets containing only one trader type. In Figure~\ref{fig:ben-results-homo-a}, we see that ZIPP traders consistently achieve significantly higher profits than ZIP traders. Figure~\ref{fig:ben-results-homo-b} presents the same results, but with results for homogeneous AA and GDX markets overlaid. Visually, profits of AA, GDX, and ZIPP fall in a similar region, with all gaining profits significantly higher than the profits attained by ZIP. AA profits are slightly, but significantly higher than GDX and ZIPP (Mann-Whitney $U$ test; $p<0.05$), while profits of GDX and ZIPP have no significant difference (Mann-Whitney $U$ test; $p>0.05$), suggesting both have similar performance.

\begin{figure*}[tb]
  \centering 
  \subcaptionbox{Balanced test ZIP:ZIPP
  \label{fig:ben-results-hetero-a}}[0.48\linewidth][c]{%
    \includegraphics[width=\linewidth]{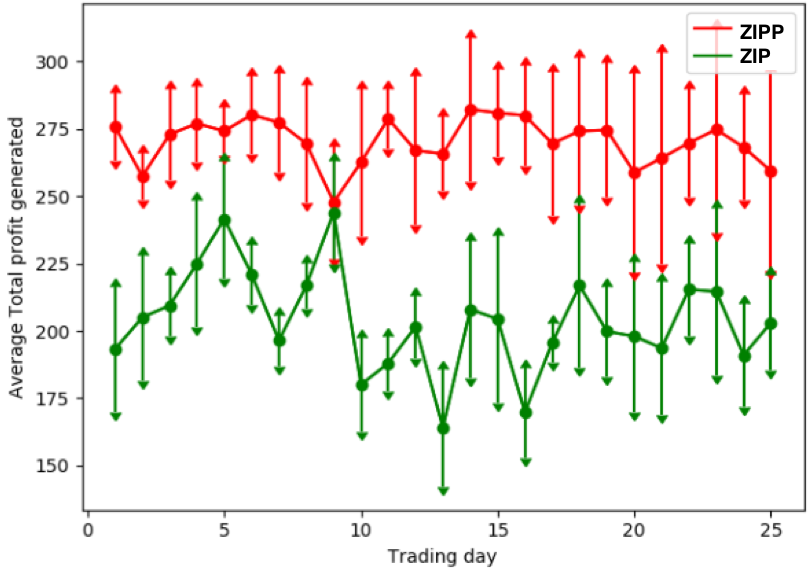}} 
    \hspace{0.01\textwidth}
  \subcaptionbox{Balanced test AA:GDX:ZIPP
  \label{fig:ben-results-hetero-b}}[0.48\linewidth][c]{%
    \includegraphics[width=\linewidth]{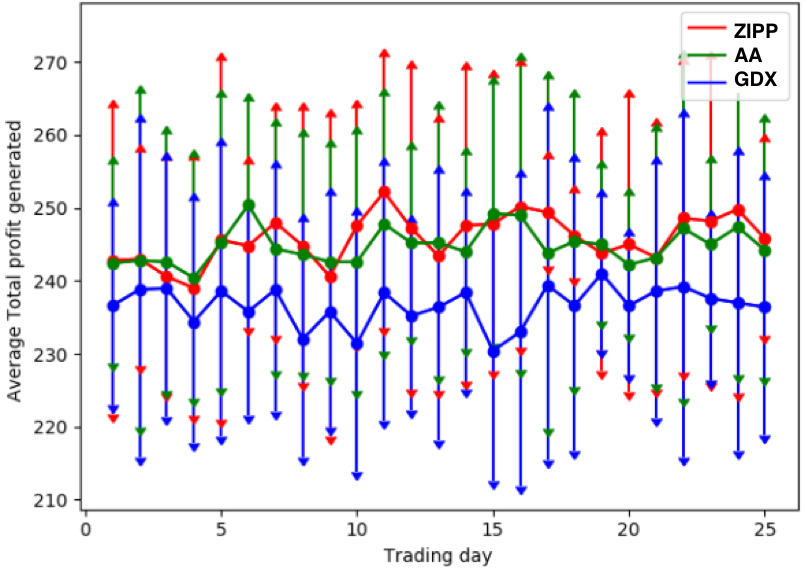}}
  \caption{Average total profit generated each trading day (heterogeneous markets).}
  \label{fig:ben-results-hetero}
\end{figure*}

\subsubsection{Heterogeneous Markets:}

Figure~\ref{fig:ben-results-hetero} shows mean profits ($\pm$ standard deviation) generated in heterogeneous markets containing multiple trader types. 
Figure~\ref{fig:ben-results-hetero-a} shows ZIPP generates significantly more profit than ZIP (Mann-Whitney $U$ test; $p<0.05$) in markets containing both agents. Markets containing AA/ZIPP and GDX/ZIPP demonstrated that AA significantly outperforms ZIPP and ZIPP significantly outperforms GDX (results not shown). 
Figure~\ref{fig:ben-results-hetero-b} shows results from a market containing three agent types: AA, GDX, and ZIPP. Here, ZIPP and AA significantly outperform GDX (Mann-Whitney $U$ test; $p<0.05$); while there is no significant difference between AA and ZIPP (Mann-Whitney $U$ test; $p>0.05$). 

Results demonstrate that adding {\em pace} to ZIP---thereby creating a ZIPP trader with a sense of urgency---increases performance; with ZIPP significantly outperforming ZIP in both homogeneous and heterogeneous markets, across a variety of market conditions. By cmparing ZIPP with other leading strategies (AA and GDX), results show that AA tends to be the most profitable, but does not always outperform ZIPP; while ZIPP outperforms GDX in heterogeneous environments. Therefore, adding urgency to ZIP results in a more profitable strategy. 

\section{\uppercase{Conclusions}}
\label{sec:conclusion}
\noindent 
We have presented an exploration of reaction speed and urgency in financial trading agents. Results demonstrate that SHVR (a simple but quick strategy) generates significantly more profit than AA (a more complex and therefore slower strategy) when reaction speed is accurately modelled. We also demonstrated that ZIPP (a ZIP trader that adjusts pace as a deadline approaches) generates significantly more profit than ZIP. These results confirm that time matters, and if we are to better understand real-world markets it is necessary for reaction speeds and urgency to be considered more thoroughly. 

Our results add to the mounting body of evidence that suggests SHVR could be a surprisingly successful trading strategy, despite being so simple. 
SHVR is the only agent considered in this paper that determines quote price solely on the basis of current order book information. Unlike AA, GDX, and ZIP, SHVR has no internal state and so does not consider {\em where} the market is trading (i.e., at what price the market is trading). This demonstrates that considering only the current order book dynamics can result in a profitable trading strategy, without considering long term price trends, or estimations of some fundamental equilibrium value. 
We therefore intend to more fully explore the use of order book metrics in trading strategies (e.g., Volume Imbalance \cite{Cartea18}; Order Book Imbalance \cite{Imaev17}) as future work. 
We note that the MAA ({\em modified} AA) strategy introduced by Cliff \cite{Cliff19} considers the order book {\em microprice}, which is a useful first step towards this goal. 

In real financial markets, order book information is known to be strategically useful, as it exposes the trading intentions of market participants. In order to hide one's intention to trade, some trading venues, described as {\em dark pools}, do not reveal market quotes and all order information remains hidden (see, e.g., \cite{MPCDark19,cryptoeprint:2020:662}, for a summary of dark pools and methods for implementing cryptographically secure dark pool mechanisms using multi-party computation (MPC)). 
This ensures no information ``leakage'', and can therefore result in a better execution price. However, orders can take much longer to execute in a dark pool as trading counterparties are more difficult to discover. Therefore, dark pools offer a trade off between immediacy and risk. Trading urgency, as we introduced in ZIPP, can help an automated strategy determine which trading venue (``lit'', for immediacy; or ``dark'', for best execution) is most appropriate to use at any given time. We intend to explore this research avenue more fully in future.

Several works have demonstrated that altering the design of trading agent experiments can raise doubt over previously established results, and how well these results translate to the real world. This has led several authors to call for more experimental {{\em realism} (e.g., \cite{CartlidgeCliff12,CartlidgeCliff18,Cliff19,DeLuca11,SnashallCliff19}). Here, we have addressed that challenge by introducing minimal models of time into the standard experimental framework. Our demonstration of significantly different outcomes when reaction times are realistically modelled adds to this growing body of evidence.

%
%
%
%

\end{document}